# Axisymmetric magnetic field effects on hollow cathode generated plasma column in APPEL-device

Y. Patil[1], S. K. Karkari[1, 2]

[1]Institute for Plasma Research, Bhat, Gandhinagar 382428, Gujarat, India,

[2]Homi Bhabha National Institute, Anushaktinagar, Mumbai 400094, India.

Email: ypatil@ipr.res.in, skakari@ipr.res.in

**Abstract**

An elongated plasma column has been successfully generated and sustained in a linear plasma device using a hollow cathode discharge in the presence of an axisymmetric magnetic field. The confinement of cold energetic electrons produced near the hollow cathode plays a crucial role in guiding the plasma along the device axis. Experimental diagnostics reveal a high concentration of energetic electrons in the peripheral region near the source, which progressively converge toward the axis at a downstream location approximately 3.0 meters from the cathode. The length of the plasma column exhibits an inverse relationship with the electron-neutral collision frequency, indicating the significance of collisional damping in the propagation of energetic electrons. These observations are further supported by fluid simulations performed using COMSOL® Multiphysics, which qualitatively reproduce the experimental trends. The results are consistent with a theoretical model previously proposed by the authors, reinforcing the understanding of energetic electron behaviour in magnetically guided plasma columns.

**Keywords:** Magnetized plasma, Hollow cathode plasma source, open magnetic system

## 1. Introduction

Low-pressure, high-density plasma sources are fundamental tools in a range of applications, including semiconductor manufacturing, spacecraft propulsion, and magnetic confinement fusion. These plasmas operate in the non-equilibrium regime, where the electron temperature remains high enough to sustain ionization, while the neutral gas remains relatively cold. Achieving high plasma densities ($\geq 10^{11}$ cm$^{-3}$) under low-pressure conditions ($\leq 10^{-3}$ mbar) requires highly efficient power coupling, effective particle confinement, and optimized gas breakdown mechanisms. Two primary classes of plasma sources are used in such regimes:

radiofrequency (RF) and direct current (DC) sources. While RF systems such as inductively coupled plasmas (ICP) [1], helicon discharges [2], and electron cyclotron resonance (ECR) plasmas [3] are widely studied for their ability to deliver high-density, uniform plasmas, DC discharges remain foundational due to their simplicity, stability, and effectiveness in low-pressure environments.

In DC plasma sources, a steady-state electric field drives a continuous flow of electrons that ionize the neutral gas. These are typically used in configurations such as DC glow discharges, magnetron sputtering sources, and cathodic arc systems. When a magnetic field is applied perpendicular to the electric field (as in crossed-field configurations), electron confinement is enhanced, resulting in greater ionization efficiency and plasma density [4]. Such magnetized DC sources have been successfully used in ion sources, negative ion beam generation, and sputtering processes [5]. Despite the operational differences between RF and DC systems, both can be configured to operate in similar pressure and density regimes. In some advanced devices, hybrid approaches are employed, where RF assisted breakdown or RF pre-ionization is followed by DC sustainment for efficient plasma maintenance [6]. Understanding the underlying energy deposition, wave particle interaction, and transport mechanisms in both RF and DC systems is critical for optimizing their performance in laboratory and industrial settings

Cold hollow cathode plasma sources have gained significant attention in recent decades due to their ability to generate high density, non-thermal plasmas under low pressure conditions. These discharges operate efficiently in the pressure range of 0.01 to 1 Torr (1.3–130 Pa), significantly lower than conventional glow discharges, which typically require 1–10 Torr (130–1300 Pa) to sustain stable plasma conditions [7].

The enhanced performance of hollow cathode discharges stems from their unique geometrical confinement, where the pendulum effect oscillation of electrons between opposing cathode walls substantially increases the effective electron path length. This leads to a higher probability of ionizing collisions and, consequently, a denser plasma [8]. The geometry of a hollow cathode plays a critical role in determining the behaviour of the plasma within it. A traditional cylindrical hollow cathode provides symmetrical confinement of electrons between opposing walls. However, when the hollow cathode is modified into a conical, hemispherical, or other non-cylindrical shapes, the spatial distribution of electric fields and the trajectories of electrons are altered. In a conical hollow cathode, for instance, the narrowing cross-section toward the tip can lead to electric field focusing, which increases the local ion and electron density near the apex of the cone [9]. This can be advantageous in applications requiring localized high-density plasmas, such as micro-thrusters, microfabrication, or nanoscale deposition [10]. When a magnetic field is applied perpendicular to the electric field (i.e., a crossed E × B configuration) inside or near a hollow cathode, it affects the charged particle dynamics. The magnetic field forces electrons to gyrate around magnetic field lines, increasing their effective path length inside the cathode cavity. This leads to longer electron residence times and more frequent collisions with neutral atoms, boosting the ionization rate and increasing plasma density [11]. Despite their long-standing history, the fundamental mechanisms governing HCD behaviour, including electron confinement, sheath dynamics, and transition modes between normal and abnormal glow, remain active areas of investigation. Moreover, advances in diagnostics and computational modelling have opened new avenues for optimizing HCD performance across diverse environments and applications.

The choice between cold and hot cathode designs influences not only the plasma initiation threshold but also the electron energy distribution function (EEDF), sheath dynamics, and overall stability of the discharge. In some advanced plasma systems, hybrid architectures are implemented such as RF-assisted DC discharges with heated cathodes to combine the benefits of both technologies, particularly for initiating breakdown under challenging vacuum conditions [12]. COMSOL Multiphysics is widely used for simulating plasmas, particularly those that are not symmetrical. The magnetic field employed in low temperature plasma devices essentially improves the plasma's characteristics and produces anisotropy, which is the source of the E × B drift [13]. Capacitively Coupled Plasma (CCP) produced by parallel plate electrodes under the influence of a transverse magnetic field influenced by E × B drift is one of our investigations. The plasma develops "S" shaped structures as a result of this open drift. A discharge occurs within the glass chamber, and a portion of the plasma cannot be probed for diagnosis. Using COMSOL Multiphysics, a three-dimensional fluid model of the magnetized CCP discharge has been created in order to examine the creation of the "S" structure. The top and bottom parallel plate electrodes are loaded with RF voltage having frequency 13.56 MHz and phase difference 180º with each other. The static magnetic field with magnitude ~ 7 mT is applied perpendicular to the discharge gape while oscillating RF electric field is along the vertical direction and resultant E × B drift occur along the length of the plate [14, 15] Recent work reports that the magnetized plasma column generated by a spiral antenna was simulated using COMSOL at an operating pressure of the $10^{-4}$ mbar and a magnetic field strength of 0.1 T, showing similar trends [16].

Recent studies carried out on the linear plasma devices and diagnostics related magnetized plasma systems further motivate the present investigation. Singh *et al.* [17] examined the influence of sheath dynamics and thermal electron populations on the resonance response of DC-biased hairpin probes, demonstrating that sheath-induced frequency shifts must be carefully accounted for in low-pressure, magnetized plasmas. Khandelwal *et al.* [18] reported the discharge properties of a magnetized cylindrical capacitively coupled plasma (CCP) and showed that the application of an axial magnetic field strongly modifies ionization dynamics, electron heating, and plasma uniformity inside cylindrical geometries. Complementary experiments by Dahiya *et al.* [19] on a geometrically asymmetric CCP configuration revealed that magnetization leads to enhanced electron confinement and pronounced axial variations in plasma potential and density. Together, these studies highlight the critical role of magnetic field topology, sheath formation, and geometry dependent electron transport in shaping plasma behaviour concepts that are directly relevant to understanding the long, magnetically guided hollow cathode plasma columns.

To address these challenges, the APPEL device (Applied Plasma Physics Experiments in Linear Device exclusively designed and established at the Institute for Plasma Research (IPR), India, to conduct physics studies related to plasma confinement, turbulence, and plasma-surface interaction studies for fusion reactors [20]. The silent feature of the APPLE device is elaborated in the next section of this article. One of the goals of the APPLE device is to test the plasma source in an axial magnetic field and study the plasma column generated by the source. The experimental and theoretical behaviour of argon and helium magnetized plasma columns are studied and subsequently presented in this article. The article further gives the COMSOL model details of the argon-helium plasma column and the plasma column length calculation using simulation results, comparison with analytical equations.

The aim of this study is to explore the behavior of hollow cathode in different values of magnetic field, and further experiments will be conducted to excel it operation for high density at low pressure. Section 2 describes the experimental set-up used for hollow cathode source studies. In section 3, the plasma characteristics of argon and helium plasma are presented. COMSOL model details with simulation results are mentioned in section 4 of the manuscript. Section 5 basically summarizes the analytical methods used to calculate the plasma column length and compare it with COMSOL simulations. Section 5 summarized the summary and conclusions of the research work.

## 2. Experimental set-up

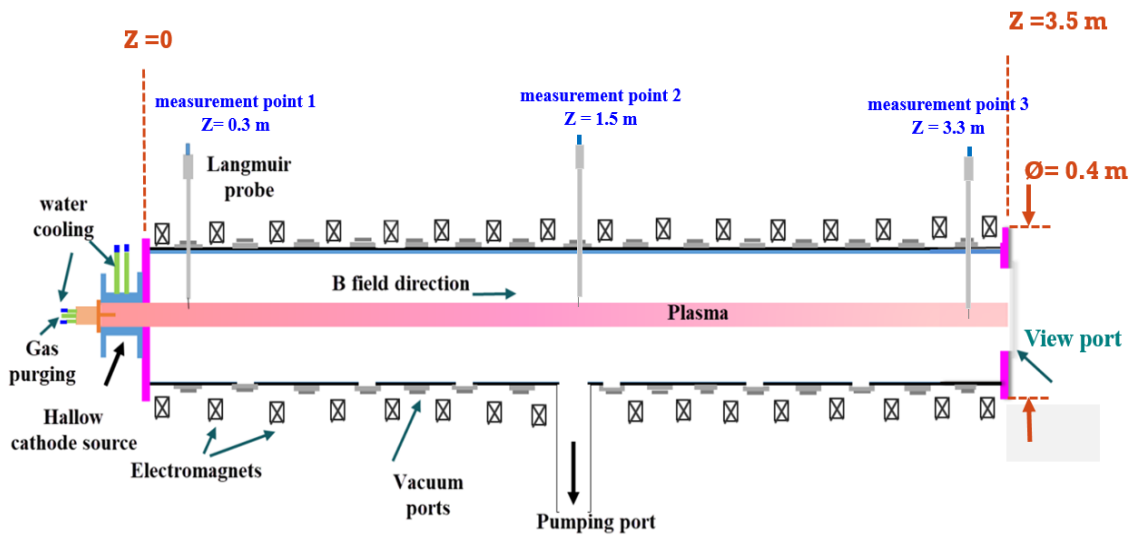

Figure 1. Schematic diagram of the APPEL device with hollow cathode source

The experiment was carried out using a linear plasma device called the APPEL (Applied Plasma Physics Experiments in Linear) device. A schematic of the setup is illustrated in Figure 1, showing a cylindrical vacuum chamber of 3.5 m length and 0.4 m diameter. The APPEL device is intended to address various basic plasma physics problems, so control plasma densities and temperatures are required. The hollow cathode source is equipped with the APPEL device located at Z = 0. As shown in Figure 1, a Langmuir probe with a diameter of 0.2 mm and a length of 0.5 cm is inserted radially at axial locations Z = 0.3 m, 1.5 m and 3 m to measure the electron density and temperature. As per schematic diagram 16 electromagnets surround the plasma chamber along its length. The axial distribution of the magnetic field along the 3.5 meter long vacuum chamber for different coil currents ranging from 130 A to 720 A. The field strength increases with higher coil currents, achieving peak values of approximately 0.05 T at 130 A and up to 0.42 T at 720 A. Such a magnetic topology is crucial for confining the plasma radially by minimizing cross-field electron diffusion. Overall, the tailored magnetic field profile plays a key role in achieving stable plasma propagation and effective confinement over long axial distances, which is essential for studies involving plasma transport and wave interactions [21].

Figure 2 presents a schematic representation of the hollow cathode source assembly. The hollow cathode source depicted is a key component of the APPEL device, designed to initiate and sustain a high-density plasma. Constructed primarily from stainless steel (SS 304), the assembly consists of two main parts: the hollow cathode and the constricted anode, both thermally and electrically isolated using Perspex flanges. Hollow cathode assembly is 10 cm long cylindrical cathode has an inner diameter of 55 mm. This cylindrical cathode structure is responsible for electron emission, which is enhanced by the hollow geometry that allows for secondary electron generation through ion bombardment. The geometry also promotes a higher degree of ionization and sustains a dense plasma within the cathode region.

Other hand Positioned coaxially a tiny tube (8.0 mm OD), which serves as the anode and is axially attached to one end of the hollow cathode. Only 5.0 mm of the anode is exposed to participate in the discharge. Constricted anode features a narrow orifice or throat that serves to stabilize the discharge and confine the arc. This constriction increases the current density and facilitates efficient ionization of the incoming gas. Anode tip precisely aligns the discharge path and ensures consistent ignition and plasma production. Argon or helium working gas is purged in the hollow cathode through the 2 mm hole present inside constricted anode.

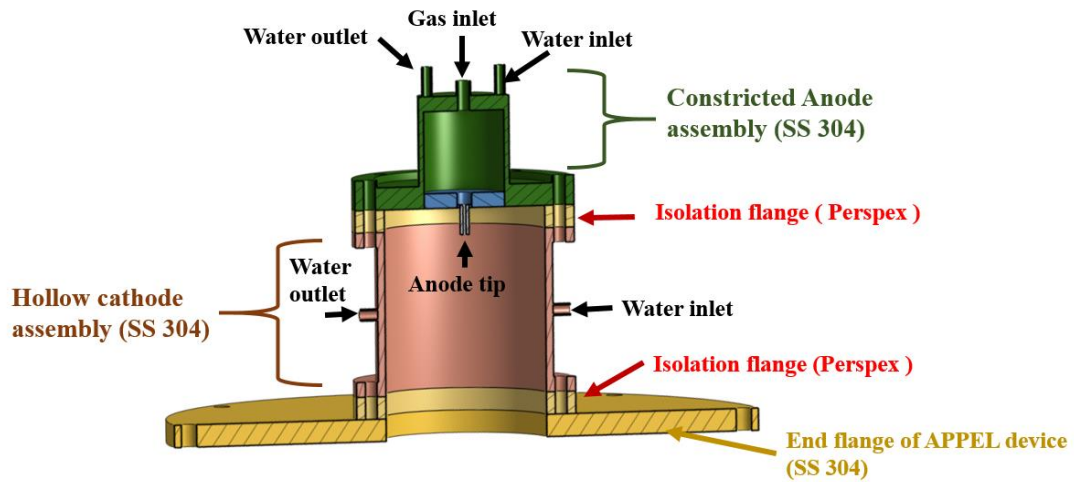

**Figure 2. Schematic diagram of hollow cathode source**

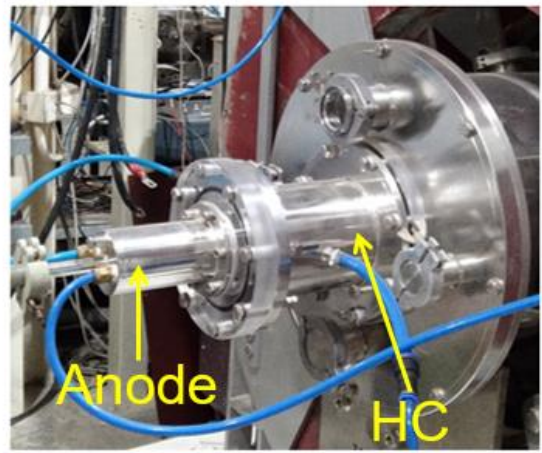

**Figure 3. photograph of hollow cathode source equipped with APPEL device**

Cooling water is circulated through designated inlets and outlets in both the anode and cathode assemblies double walled chambers to manage thermal loads during operation, thereby maintaining stable performance and prolonging component life. The double-wall water cooling chamber provides forced water cooling to the cathode and anode. Both electrodes are isolated from the grounded chamber.

Figure 3 shows the photograph of the hollow cathode (HC) plasma source mounted on the vacuum chamber flange of device. Figure 4 illustrates the electrical circuit configuration employed to operate the hollow cathode plasma source.

The circuit is powered by a variable DC voltage 'Vdc' using DC power supply (Make: Magna-power, model: TSD 2000-10/415) to sustain discharge between the anode and the hollow cathode. A load resistor 'RL' is connected in series with the power supply to regulate the discharge current and acts as a current limiting resistor protects the discharge from switching to an arc mode and protecting the power supply and electrodes from overcurrent damage. Hollow cathode and constricted anode is grounded via resistance 5kΩ and 500 Ω to form a potential divider circuit, allowing for fine control of the bias voltage applied across the plasma. Discharge current measured in terms of voltage using resistance 'Rdis'.

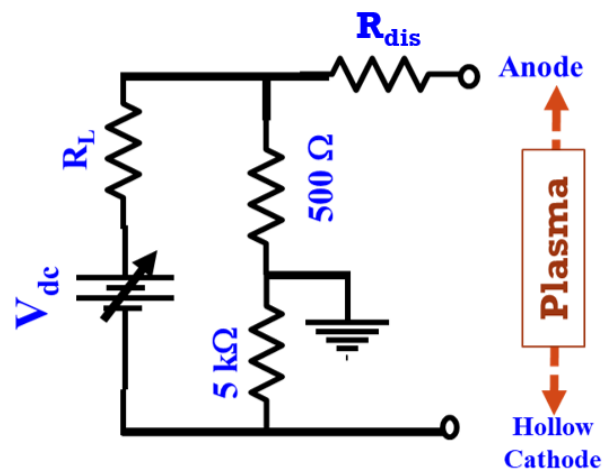

Figure 4. The electrical circuit of the hollow cathode source

The hollow cathodes typically operate on the left-hand branch of the Panchen's curve. They have relatively greater efficiency than planar cathodes because their geometry enables for the trapping of secondary electrons inside the hollow cavity. Constriction of the anode further enhances the discharge efficiency since the current densities (Je ~50 A/m$^2$) at the anode are extremely high. Additionally, differential pumping due to the flow of gas through the anode gives rise to high pressure behind the tip of the anode. As a result, it reduces the E/p ratio which helps in the formation of the discharge at a low pressure inside the main chamber [22]. Present work is focused to study the hollow cathode source performance in the axial magnetic field and it can be used as cost-effective source for plasma wall interaction studies. A helium plasma column extending over a length of 3.5 m is observed during the application of a magnetic field of 20 - 50 mT. The axisymmetric magnetic field lines produced in the APPEL device are tangential to the surface of the hollow cathode, while the magnetic field points directly toward the anode. A long plasma column is formed as a result of the applied magnetic field. To understand the behaviour of the long plasma column and the underlying physics, experimental studies have been conducted and are interpreted in the next section. The plasma parameters along the axial direction measured using three Langmuir probes were installed at axial positions of 0.3m, 1.5m, and 3m from the plasma source. Each probe is consists of cylindrical metallic tip with diameter of the order of a few microns and length of 5 cm ensuring minimal perturbation to the plasma while providing reliable current voltage characteristics. The probes were aligned parallel to the plasma flow to reduce sheath distortion and ensure accurate measurements of electron temperature, plasma potential and electron density at various locations along the axis.

3. Plasma characteristic of Argon and helium plasma

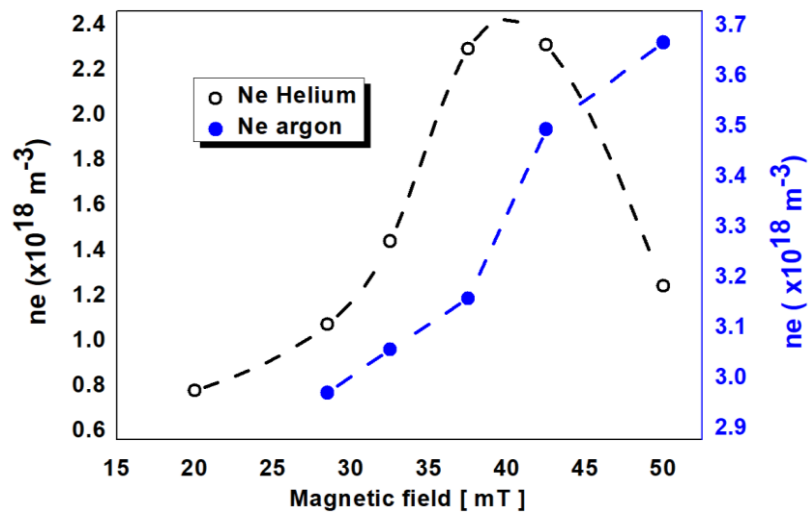

**Figure 5 a. Comparison of electron density of Helium and argon plasma with respect to magnetic field pressure 2 Pa at axial distance 0.3 m.**

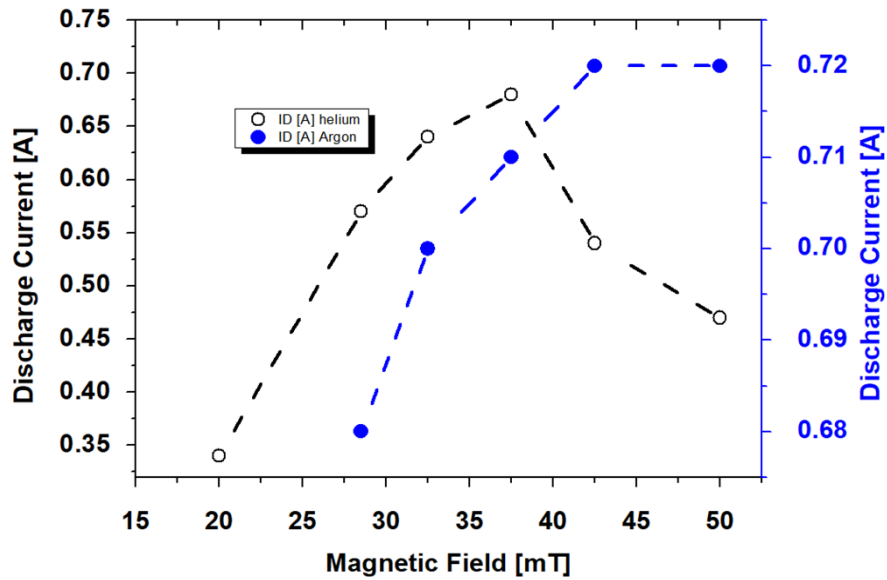

**Figure 5 b.** Comparison of discharge current of Helium and argon plasma with respect to magnetic field pressure 2 Pa at axial distance 0.3m.

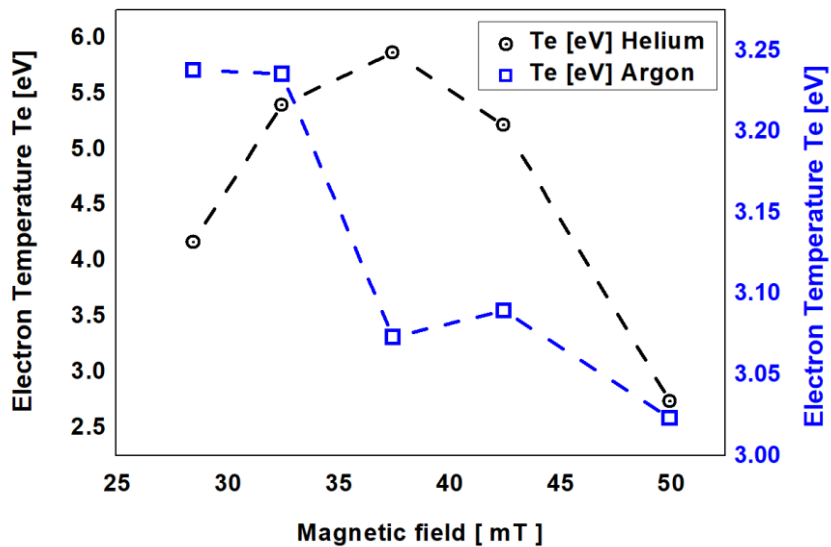

**Figure 5 c.** Comparison of Electron temperature of Helium and argon plasma with respect to magnetic field pressure 2 Pa at axial distance 0.3 m.

Comparative studies of argon and helium plasmas were performed under identical operating conditions, with constant DC power and a fixed pressure of 2 Pa. The helium plasma consistently formed a stable and extended column approximately 3.5 meters in length across the entire range of applied magnetic field strengths. In contrast, the length of the argon plasma exhibited significant variation with changes in the magnetic field. Figure 5 presents a comparison of the discharge parameters for helium and argon plasmas as a function of magnetic field strength, measured at an axial position of 0.3 m. As the axial magnetic field strength increases, Figure 5 (a) and (b) demonstrate that both plasma density and discharge current first increase before gradually decreasing as the magnetic field strength beyond a threshold value. It appears that argon has a greater threshold value than helium. Below the threshold value, the secondary electrons are confined in circular orbits inside the hollow cathode, increasing the discharge current and causing more ionizing collisions. However, the secondary electrons have a propensity to strongly magnetize, making it impossible for them to emanate from the cathode surface at magnetic fields over the threshold level. In other words, as the magnetic field increases, there are three effects that occur: (1) a decrease in discharge current; (2) a decrease in secondary electron yield caused by the re-capture of electrons at the hollow cathode surface; and (3) an increase in ion bombardment at the cathode surface. Due to the considerable reduction in power dissipated on the cathode walls as a result, helium has a higher plasma density than argon for a given discharge power.

At magnetic field higher than the threshold value, the positive helium ions are magnetized whereas argon ions are not magnetized at all the values of magnetic fields. Because of this, the efficiency of secondary electron emission from helium ion bombardment at the cathode rapidly declines as the magnetic field increases. The energy loss via ions at the cathode sheath is comparably lower for helium ions than argon. The partial magnetization of helium ions is another potential explanation for the increase in helium plasma density. The 3.5 m long helium plasma column is found to be stable, collimated, and effective under a variety of magnetic fields. This is due to the helium being radially confined by the axial magnetic field.

Figure 5. c. illustrates the variation of electron temperature ($T_e$) as a function of magnetic field for helium and argon plasmas. A clear distinction is observed in the magnetic field dependence of $T_e$ for the two gases. In helium, $T_e$ increases sharply from approximately 4.2 eV at 30 mT to a peak value of ~ 5.9 eV around 37 mT, followed by a gradual decline with further increase in magnetic field. This trend indicates enhanced electron confinement and heating efficiency due to the magnetic field, which reduces radial diffusion and increases the effective electron path length, thereby promoting more ionizing collisions. In contrast, argon exhibits a relatively flat $T_e$ profile, with a slight decrease from 3.25 eV to 3.02 eV over the same magnetic field range.

This behaviour can be understood through electron dynamics and collisional processes in DC magnetized plasma. Calculations show that although the electron larmor radius is slightly smaller for argon (~1.01 mm) than helium (~1.28 mm) at 35 mT, the lower ionization energy of argon (15.8 eV) compared to helium (24.6 eV) means that argon achieves sufficient ionization even at lower Te. Furthermore, helium's lower electron-neutral collision frequency ($v_{en} \sim 10^7$ s$^{-1}$) allows the magnetic field to dominate electron motion ($\omega_{ce} \gg v_{en}$), leading to strong magnetization and enhanced confinement. In contrast, argon's higher collisionality ($v_{en} \sim 5 \times 10^8$ s$^{-1}$) suppresses the magnetic field's ability to confine electrons effectively ($\omega_{ce} \lesssim v_{en}$), limiting the increase in Te. As a result, the magnetic field enhances electron confinement and ionization more effectively in helium than in argon under similar operating conditions.

In the absence of an axial magnetic field, the secondary electrons generated by positive ion bombardment at the cathode surface are reflected from the opposing cathode sheaths inside the hollow cathode. These secondary electrons experience a considerable resistance when trying to traverse the magnetic field and reach the anode when a magnetic field is applied. They can only be elastically distributed into the core region after travelling a sufficient axial distance along the magnetic field lines.

Figure 6 shows the radial hot electron population distribution in helium plasma column near the source at distance 0.3 m and at 3 m distance always from the source and it showed that at the axial distance 0.3 m the density of hot electrons is at periphery region while at 3 meter the hot electrons density is higher at the centre of plasma column.

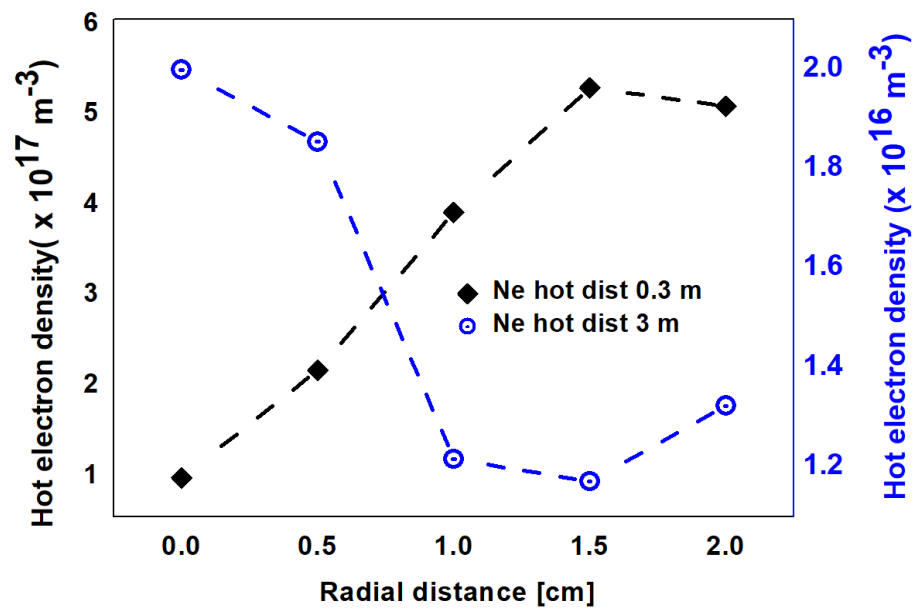

**Figure 6. Radial distribution of hot electron density for helium at 0.3 m and 3.0 m.**

Figure 6 clearly indicates that hot electrons travel long distance to come in centre of helium plasma column as reported above. The scattered electrons eventually reach the anode in the core and constitutes the discharge current. The electron-ion pair in the positive column is created ohmically i.e. I²R heating and extends far enough along the magnetic field from the anode. A schematic illustration of this mechanism represented in figure 7.

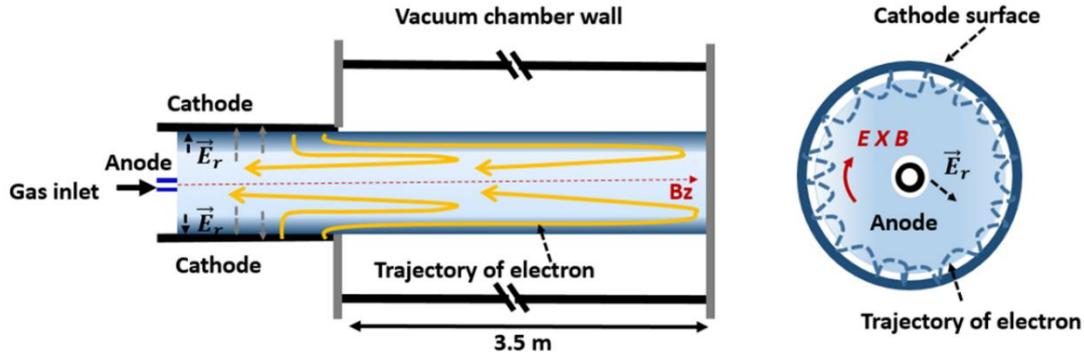

**Figure 7. Schematic diagram of hollow cathode discharge mechanish occure in APPEL linear device**

To further extend the understanding of the helium plasma column, Table 1 summarizes the axial plasma parameters measured along the discharge column at B = 50 mT. The table lists the experimentally obtained values of plasma potential, electron density, and electron temperature at representative axial locations. These data clearly illustrate the evolution of plasma characteristics from the source region toward the downstream end

**Table 1 Axial parametrs of helium Plasma colum at field 50mT**

| Axial location [m] | Plasma potential (eV) | Electron density ($m^{-3}$) | Electron temperature (eV) |
|---|---|---|---|
| 0.3 | 10.3 | $2.9 \times 10^{18}$ | 5.2 |
| 1.5 | 4.5 | $7 \times 10^{17}$ | 4 |
| 3 | 3.2 | $6.9 \times 10^{17}$ | 11.5 |

In the case of helium plasma at an applied magnetic field of 50 mT, a well formed and extended plasma column was observed along the axial length of the APPEL device. The experimental measurements show that the plasma potential decreases from approximately 10.3 eV near the source region (z = 0.3 m) to 3.2 eV at the downstream end (z = 3.0 m), indicating a smooth potential drop along the magnetic field lines that aids axial electron transport. The electron density decreases from about $2.9 \times 10^{18}$ m⁻³ near the source to $6.9 \times 10^{17}$ m⁻³ downstream, demonstrating efficient confinement and sustained ionization along the magnetic field. The electron temperature ($T_e$) varies significantly with axial position, showing a moderate value of ~5.2 eV near the source, a slightly cooler region of ~4 eV in the mid-plane (z = 1.5 m), and a noticeable increase up to ~11.5 eV toward the far end of the plasma column.

This rise in Te at the downstream location suggests the presence of energetic electrons transported along the magnetic field, possibly due to reduced collisional damping and enhanced parallel acceleration mechanisms.

## 4. COMSOL Model description

To understand the plasma behaviour in the hollow cathode discharge under an axial magnetic field, a two-dimensional (2D) axisymmetric model was developed using COMSOL Multiphysics ® 6.0. The simulation focuses on capturing the essential physical processes that determine plasma generation, confinement, and transport. In this study, the long-magnetized plasma column was sustained in the APPEL device by E × B drift. The aim of this study is to calculate the length of the plasma column and validate the analytical method.

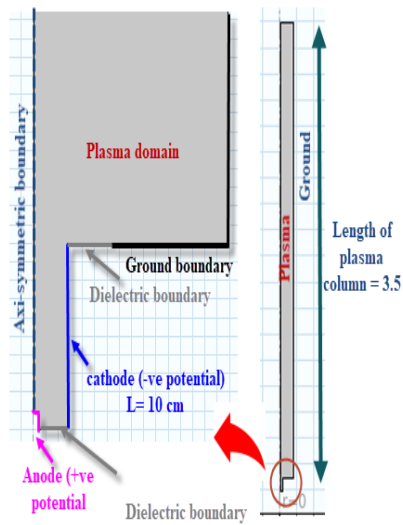

**Figure 8. Geometry and schematic of the model**

**Table 2: Reaction cross section of Argon plasma**

| No. | Reaction | Δε [eV] |
|---|---|---|
| 1. | $e + Ar \rightarrow Ar^+ + 2e$ | 15.7 |
| 2 | $e + Ar \rightarrow Ar^* + e$ | 11.56 |
| 3 | $e + Ar^* \rightarrow Ar^+ + 2e$ | 4.14 |
| 4 | $e\,(slow) + Ar^* \rightarrow Ar + e(fast)$ | -11.56 |
| 5 | $Ar^* + Ar^* \rightarrow Ar^+ + Ar + e$ | |
| 6 | $Ar^* + Ar \rightarrow 2Ar$ | |
| 7 | $Ar^* \rightarrow Ar$ | At plasma boundary |
| 8 | $Ar^+ \rightarrow Ar$ | At plasma boundary |

| No. | Reaction | Δε [eV] |
|-----|----------|---------|
| 1.  | $e + He \rightarrow He^+ + e$ |  |
| 2   | $e + He \rightarrow He^* + e$ | 19.8 |
| 3   | $e + He \rightarrow He^+ + 2e$ | 24.6 |
| 4   | $He^+ \rightarrow He$ | At plasma boundary |
| 5   | $He^* \rightarrow He$ | At plasma boundary |

**Table 3: Reaction cross section of Helium plasma**

In this work, a 2-dimensional (2D) axisymmetric plasma simulation was performed and cylindrical co-ordinate system considered for model with co-ordinates r, z . Magnetic Field (mf), Plasma (plas) module of comsol were used. Simulations carried out in two steps where static magnetic field of APPEL device calculated using magnetic field module of COMSOL software.

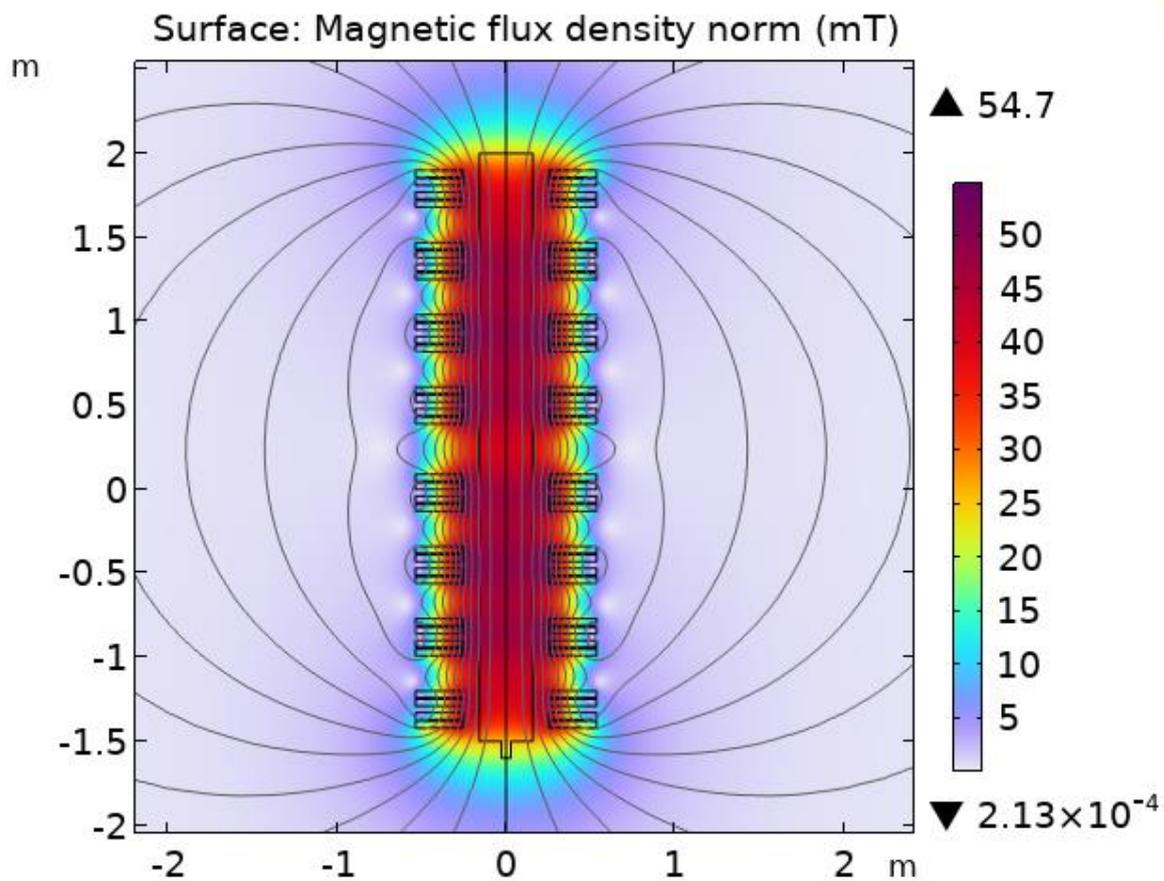

**Figure 9. 2D plot of magnetic filed in APPEL device generated by COMSOL simulation for operating current 100 A**

Plasma density and temperature calculated using plasma module of the COMSOL. 2D axisymmetric geometry with boundary condition used for the simulation was illustrated in figure 8. As per figure 8 the constricted anode mentioned as anode is asigned positive potential and hollow cathode asigned negative potential. Both electrodes are separated from each other as well as wall of chamber by dielectric boundaries. 3.5 m long cylindrical chamber wall asigned to be ground. COMSOL solves the drift-diffusion equations to compute the electron density (ne) and mean electron enegies as well as use the supporting electromagnetic equations to compute corresponding fields and are reported by many researchers in research articles [23, 24, 25].

COMSOL Multiphysics employs the drift-diffusion approximation in its Plasma Module to efficiently model low temperature, quasi neutral plasmas. This approach simplifies the complex kinetic equations governing plasma behavior into a set of macroscopic equations, balancing computational efficiency and physical accuracy. This method is particularly effective for scenarios where the electron mean free path is small compared to the system size, and the plasma is in a steady state. Solving the full momentum equations for each species, including ions and neutrals, can be computationally intensive, while in low temperature plasmas the electron energy distribution often deviates from Maxwellian distributions, and the drift velocity is small compared to thermal velocities. Under these conditions, the drift-diffusion approximation provides a good balance between accuracy and computational feasibility.

Two types of plasma, argon and helium, are considered in the simulations. The relevant interactions and their energy thresholds are catalogued in Tables 2 and 3 [26], which outline the reaction cross-sections for argon and helium plasma, respectively. For argon, key reactions include the formation of metastable states and ionization, while for helium, reactions involve direct ionization and neutralization at the plasma boundary. Gas pressure is assumed to be uniform in model geometry and is 2 Pa for both Argon and helium. The software also provides the choice of the electron energy distribution function (EEDF) can be defined by Maxwellian, Druyvesteyn, and Generalized distribution functions. In this studies, the Druyvesteyn EEDF is used. Magnetic field (B) is calculated using the magnetic field module of the software and parametrically varied from B = 0 to B = 50 mT. As reported above the plasma is treated as a multiple fluid with many charge species in the COMSOL simulation. It determines each species, electric field, momentum, and energy by solving the Poisson, continuity, and drift-diffusion equations. The electrons' continuity equation (e), Metastable atoms (*) and positive ions (i) can be expressed as follows:

Electron Continuity Equation

$$\frac{\partial n_e}{\partial t} + \nabla \cdot (\Gamma_e) = S_e \qquad (1)$$

Ion Continuity Equation

$$\frac{\partial n_i}{\partial t} + \nabla \cdot (\Gamma_i) = S_i \qquad (2)$$

Metastable Continuity Equation

$$\frac{\partial n_*}{\partial t} + \nabla \cdot (\Gamma_*) = S_* \qquad (3)$$

Where, Γ represents particle flux and S denotes the source term defined in the simulation. The source term $S_k$ captures interactions among species

$$S_k = \sum_{a,b} K_{ab} n_a n_b \quad (4)$$

Where $K_{ab}$ is the collision reaction rate coefficient, which is given as $K_{ab} = \int f(\varepsilon)\, \sigma_{ab}(\varepsilon)\, d\epsilon$, and $n_a$ and $n_b$ are the densities of the colliding particles. The electron velocity is represented by $u$, the collision cross section by $\sigma_{ab}$, and the electron energy distribution function by $f(\varepsilon)$. $K_{ab}$ is calculated using Electron energy distribution function (EEDF) and collision cross-sections. The drift-diffusion approximation provides the following expression for the momentum balance for the charge particles:

$$\Gamma_e = -n_e \mu_e E - D_e \nabla n_e \quad (5)$$

$$\Gamma_i = Z n_i \mu_i E - D_i \nabla n_i \quad (6)$$

$$\Gamma_* = -D_* \nabla n_* \quad (7)$$

Here, μ, D, E are the particle species mobility, diffusion coefficient, and DC electric field respectively. The energy balance equation is solved for electron only to obtain electron temperature.

$$\frac{\partial n_\varepsilon}{\partial t} + \nabla \Gamma_\varepsilon = -\Gamma_\varepsilon \cdot E - S_\varepsilon \quad (8)$$

Here, $\Gamma_\varepsilon$  $\Gamma_\varepsilon = -n_e \varepsilon \mu E - D_\varepsilon \nabla n_\varepsilon$ where, mean energy flux $n_\varepsilon = n_e \bar{\varepsilon} = n_e \cdot \left(\frac{3}{2} T_e\right)$

Where nε, ε, $T_e$, $\Gamma_\varepsilon$, $S_\varepsilon$ are the electron energy density, electron mean energy, electron temperature, Electron energy flux, electron energy source, respectively. Poisson's equation is solved concurrently with the fluid equations to determine the self-consistent electric field:

$$\nabla \cdot \vec{E} = \frac{e}{\epsilon_0}(n_i - n_e) \quad (9)$$

$\vec{E} = -\nabla \varphi$ Where, $\varphi$ is the electrostatic potential and $\epsilon_0$ is the vacuum permittivity. The coupled equations are numerically solved using the finite element method in COMSOL using initial seeding parameters and boundary conditions. While solving the model equation, the plasma is assumed to be weakly ionized and having a uniform pressure throughout the discharge. The ions are considered to be at room temperature. The electrons are magnetized under this field and therefore the diffusion coefficient and mobility of electrons in the model equation is replaced by the tensor quantities. As reported above the magnetic field used to calculate the tensor properties directly import from magnetic field module of the COMSOL as per equation written below. Magnetic field interaction with plasma is considered through change in mobility tensors. This tensor is replaced by scalar mobility in the continuity equations.

$$\mu_{magnetized}^{-1} = \begin{bmatrix} \mu_e^{-1} & -B_z & B_y \\ B_z & \mu_e^{-1} & -B_x \\ -B_y & -B_x & \mu_e^{-1} \end{bmatrix} \begin{bmatrix} \mu_{xx} & \mu_{xy} & \mu_{xz} \\ \mu_{yx} & \mu_{yy} & \mu_{yz} \\ \mu_{zx} & \mu_{zy} & \mu_{zz} \end{bmatrix}$$

All the ground boundaries are treated at zero potential (V = 0). The boundary condition for the electron continuity and energy balance equation is given by (10) and (11) which assume that the electron particle and energy losses are controlled by the electron thermal flux towards the plates [27].

$$\vec{n} \cdot \Gamma_e = \frac{1}{2} v_{th} n_e \qquad (10)$$

$$\vec{n} \cdot \Gamma_\varepsilon = \frac{5}{6} v_{th} n_\varepsilon \qquad (11)$$

Where, $v_{th} = \sqrt{\frac{8K_b T_e}{\pi m_e}}$ is the thermal velocity of electrons. In this operating range, the secondary electron emissions can be considered and is 0.2 [28, 29]. The initial seed parameters to initiate the simulation are specified in Table 4.

**Table 4 Initial seeding parameters used for simulation**

| Parameter | Initial seed values |
|---|---|
| Electron density | $10^{15}$ m$^{-3}$ |
| Electric potential | 4 V |
| Electron temperature | 4 eV |
| Ion temperature | 0.026 eV |
| Time step of solver | $10^{-8}$ s |

The 2D electron density plot at different magnetic field values up to 50 mT is used to calculate the length of the plasma column and as shown in figure 10. Since, APPLE plasma chamber is 3.5 m and helium is generating long plasma column lengths are > 3.5 m. Therefore, helium plasma column at 50 mT reported in figure 10. The corresponding Electron temperature at different values of magnetic field up to 50 mT reported in figure 11. Figure 12 and 13 show the comparison of electron density and temperature calculated using COMSOL and experimental results for argon plasma.

5. **Simulation results**

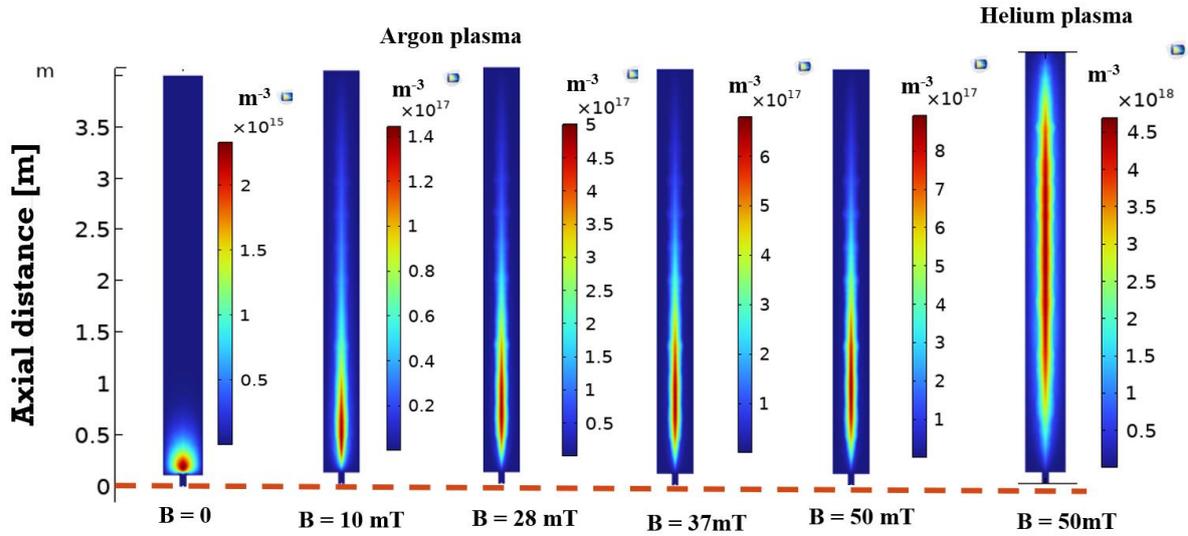

**Figure 10. Axial plasma density distribution of argon and helium plasmas under varying axial magnetic fields in a hollow cathode discharge configuration.**

Figure 10 presents the simulated axial plasma density distributions for both Argon and Helium plasmas under varying magnetic field strengths ranging from 0 mT to 50 mT. The application of an axial magnetic field significantly alters the spatial confinement and distribution of the plasma along the discharge channel. In the case of argon plasma, a low-density and short-length plasma plume is observed in the absence of a magnetic field (B = 0 mT), confined to within a few centimeters near the cathode. As the magnetic field is increased (B = 10 – 37 mT), the plasma becomes more elongated and concentrated along the axis, with peak densities reaching up to ~ $6 \times 10^{17}$ m$^{-3}$. The collimation and axial extension of the plasma are primarily due to the enhanced confinement of electrons, which limits their radial diffusion and increases their path length for ionizing collisions. At B = 50 mT, a saturation behavior in axial extension is evident, indicating that further confinement does not significantly increase the axial plasma reach. In contrast, the helium plasma at B = 50 mT exhibits a much higher peak density (~ $4.5 \times 10^{18}$ m$^{-3}$) and an extended axial distribution compared to argon. This can be attributed to the lighter mass and higher thermal velocity of helium electrons, which, when confined by the magnetic field, lead to more efficient ionization over longer axial distances. The more pronounced confinement and penetration of helium plasma may also result from the lower collisional cross-sections and higher electron mobility, allowing electrons to maintain sufficient energy for ionization even far from the cathode. These results clearly demonstrate the role of axial magnetic fields in tailoring the discharge characteristics of hollow cathode plasmas. While both gases benefit from magnetic confinement, helium exhibits superior axial extension and density under the same magnetic conditions, making it a favorable candidate for long-range plasma applications such as beam neutralization, material processing, or space propulsion systems.

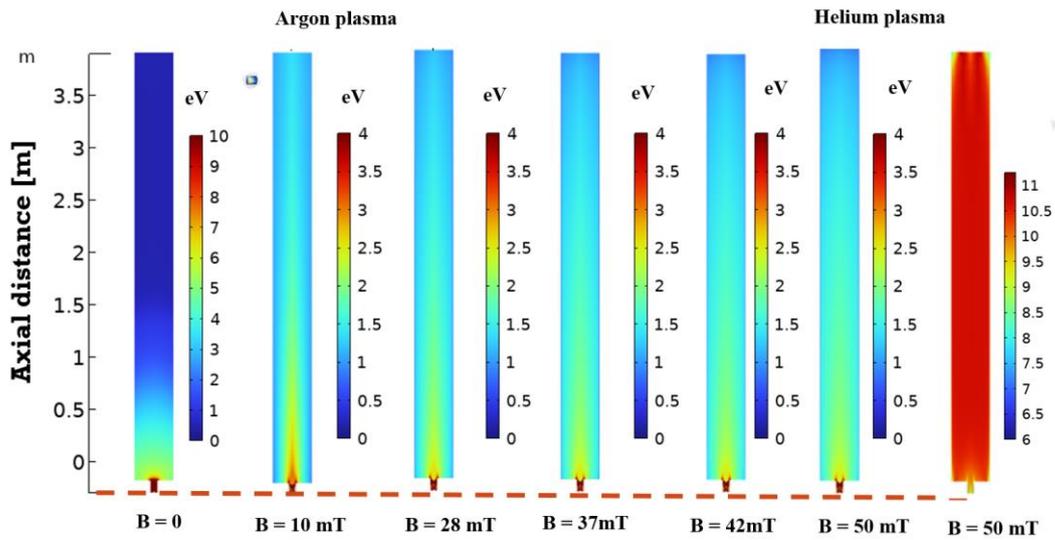

**Figure 11. Axial distribution of electron temperature in argon and helium plasmas under varying axial magnetic fields in a hollow cathode discharge.**

Figure 11 illustrates the axial distribution of electron temperature (in eV) for both argon and helium plasmas under different axial magnetic field strengths within the hollow cathode discharge configuration. In the absence of an external magnetic field (B = 0), argon plasma exhibits a localized region of high electron temperature near the cathode orifice, reaching up to ~10 eV. However, with the application of an axial magnetic field (B ≥ 10 mT), the temperature distribution elongates along the axial direction while significantly reducing the peak temperature (down to ~3 – 4 eV), indicating better confinement and reduced energy loss via diffusion. As the magnetic field increases from 10 mT to 50 mT, the electron temperature in argon plasma remains relatively steady along the axis, suggesting the establishment of a stable confinement regime. Helium plasma, in contrast, shows a substantially higher and more uniform electron temperature along the axial direction at B = 50 mT, with values ranging between 6 and 11 eV. This is attributed to helium's higher ionization potential and lighter mass, which allows more efficient acceleration and energy retention in the presence of magnetic confinement. Overall, the enhancement in electron temperature distribution with increasing magnetic field strength signifies improved plasma confinement and energy coupling. Helium plasma demonstrates superior energy confinement compared to argon, highlighting its potential advantages for applications requiring higher electron energies and extended plasma columns.

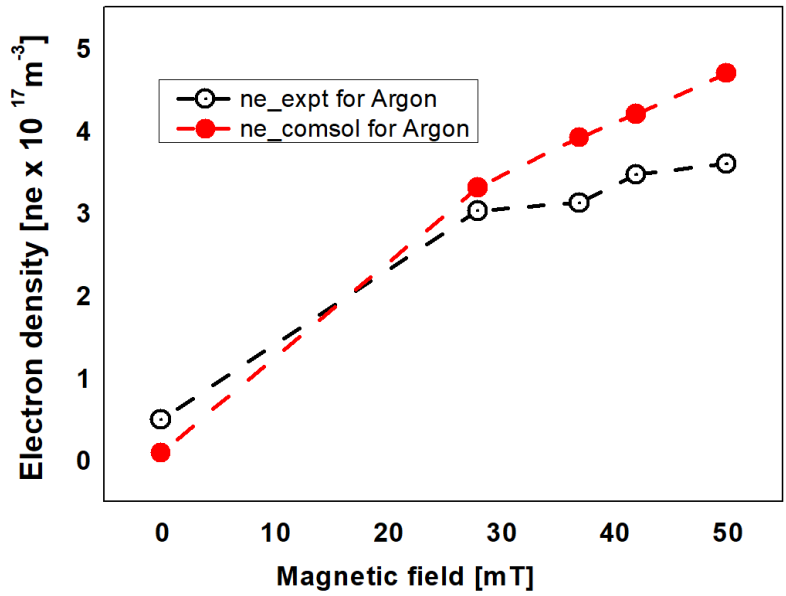

**Figure 12.** Comparison of experimental measured electron density with COMSOL simulation for various values of magnetic field

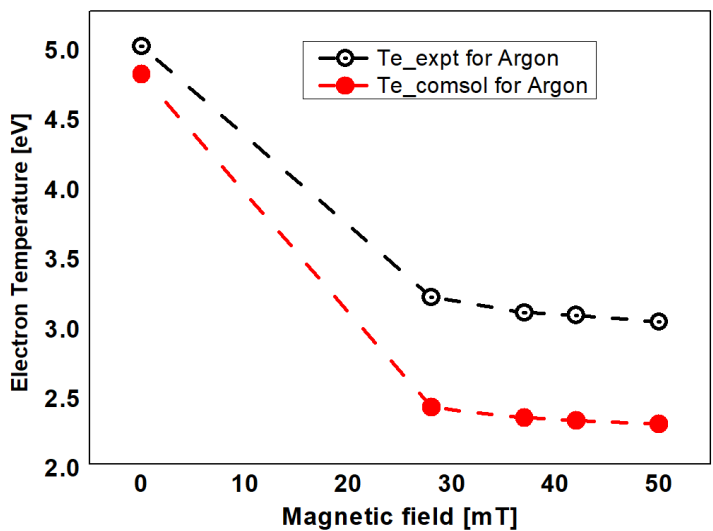

**Figure 13.** Comparison of experimental measured electron temperature with COMSOL simulation for various values of magnetic field

With magnetic field the electron density increases and electron temperature decerases. At lower magnetic fields (0 – 20 mT), both experimental and simulation values are quite close. As the magnetic field increases beyond 30 mT, simulation predicts higher electron density than measured experimentally. Real plasma may experience enhanced transport, wall losses, or instabilities not fully captured in the COMSOL model. Electron-neutral or electron-ion collisions might lead to losses not well represented in simulation. Red portion of figure 10 is used to estimate the length of the plasma coulmn. The magnetic field positively influences electron density of plasma. larmor radius decreases from 1 mm to 0.3 mm while Bohm Diffusion coefficient decreases and confinement increases. For B = 50 mT, larmor radius $r_e = \frac{m_e v_{th}}{eB}$ is 0.2 mm Bohm diffusion coefficient $D_B = \frac{1}{16}\frac{kT_e}{eB}$ is 4 m²/s.

## Analytical Estimation of Plasma Column Length

The axial length of a magnetized plasma column can be analytically estimated by balancing the electron fluxes along and across the magnetic field lines, as shown schematically in Figure 14. Under steady-state conditions, the total current entering and leaving the plasma column must remain conserved.

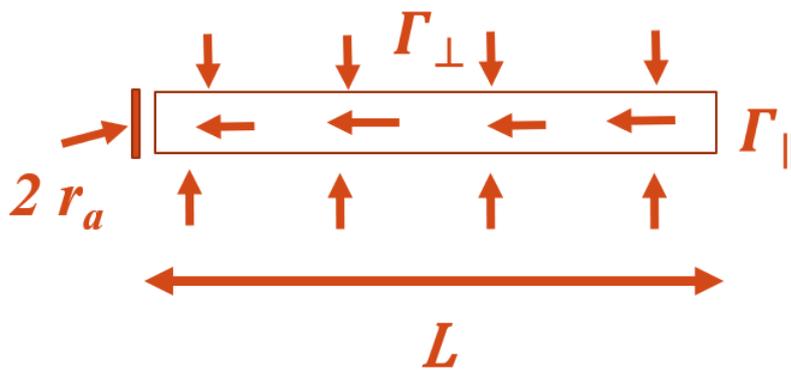

Figure 14. Schematic diagram of electron flux in the long plasma column

Accordingly, the parallel electron flux $\Gamma_\parallel$, directed along the axis and collected by the anode, must be balanced by the radial (perpendicular) flux $\Gamma_\perp$ leaving the cylindrical surface of the column. $r_a$ denote the anode radius and L the length of the hollow cathode region. Assuming cylindrical symmetry, the electron fluxes can be expressed as:

$$(2\pi\, r_a L)\Gamma_\perp e = (\pi r_a^2)\, e\Gamma_\parallel \qquad (12)$$

The perpendicular particle flux $\Gamma_\perp$ in a magnetized plasma can be described by Fick's law of diffusion as where D is the diffusion coefficient across the magnetic field and $\frac{\partial n}{\partial r}$ is the radial density gradient. In the presence of a magnetic field, charged particles undergo cyclotron motion, which inhibits their motion perpendicular to the field lines. This effect modifies the diffusion coefficient which is classical cross-field diffusion coefficient in magnetized plasma equation 13 repesent it. This expression highlights how collisions and magnetic confinement together determine the rate of plasma particle transport perpendicular to the magnetic field.

$$\Gamma_\perp = -D\frac{\partial n}{\partial r} \approx -\frac{kT_e}{m\vartheta_m(1+\omega_{ce}^2\tau_m^2)}\left(\frac{n_o}{r_a}\right) \qquad (13)$$

The particle flux parallel to the magnetic field, $\Gamma_\parallel$ describes how charged particles move along the field lines under the influence of an electric field or density gradient. Equation 14 gives the flux along magnetic field.

$$\Gamma_\parallel = -n\mu_\parallel E_z \approx -\frac{kT_e}{m\vartheta_m}\frac{\partial n}{\partial z} \approx \frac{kT_e}{m\vartheta_m}\left(\frac{n_0}{L}\right) \qquad (14)$$

Using equation 12, 13 and 14 we get the equation for length of magnetized plasma coloumn L.

$$L \approx \frac{\omega_{ce}}{\sqrt{2}\,v_m} r_a \qquad (15)$$

Alternatively, the plasma column can be modeled as a long cylindrical conductor with electrical conductivity $\sigma_{dc}$ The effective resistance of $R_p$ the plasma is given by [30,31].

$$\frac{L}{A} \approx R_p\, \sigma_{dc} \qquad (16)$$

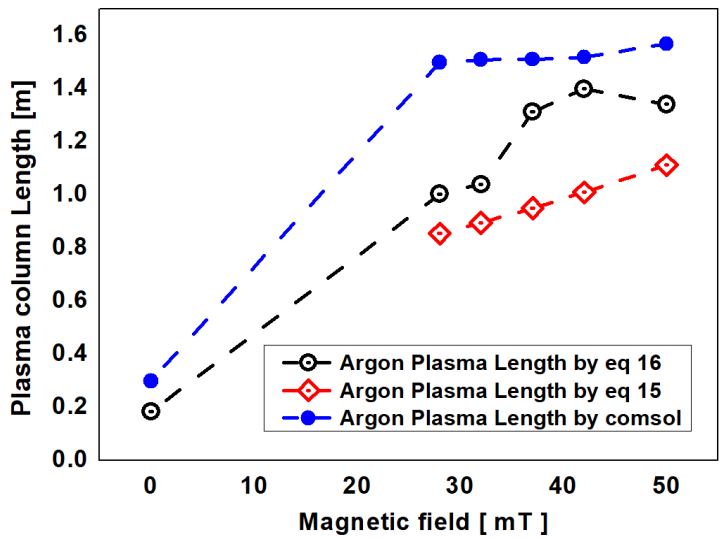

**Figure 15.** Plot of the plasma coulmn length with respect to axial magentic field comparison between analytical method and comsol simulation for Argon plasma

As per figure 15 the magnetic field increases, the Argon plasma column length increases for all methods up to a certain point (saturation). This is consistent with the physical expectation: increased magnetic confinement enhances axial transport and reduces perpendicular diffusion, allowing the plasma to extend further along the device. Equation 15 Consistently gives the shortest plasma length prediction while the equation 16 and COMSOL simulation shows same trends for argon plasma.

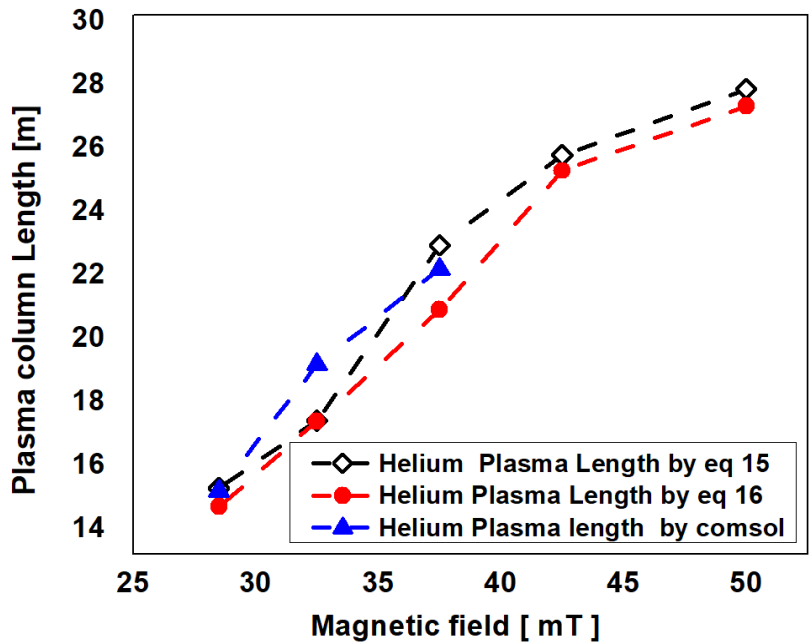

**Figure 16. Plot of the plasma coulmn length with respect to axial magentic field comparison between analytical method and comsol simulation for Helium plasma**

Figure 16 gives the Plasma column length for helium plasma. The plasma length was evaluated using two analytical approaches (Eq. 15 and Eq. 16), as well as numerical simulation results obtained via COMSOL Multiphysics. The data demonstrate a clear increasing trend in plasma column length with rising magnetic field using all methods. This behavior aligns with theoretical expectations, where stronger magnetic confinement improves plasma stability and elongates the plasma column. Quantitatively, at 30 mT, the plasma column lengths predicted by Eq. 15, Eq. 16, and COMSOL are approximately 15.2 m, 14.8 m, and 15.1 m, respectively for helium. At the upper limit of 50 mT, the respective values are 28.0 m, 27.3 m by Eq. 15 and Eq. 16. while COMSOL simulation have limitation therefore, results are not reported for higher values of magnetic field.

## 6. Summary

This In this work, we investigated the effects of axial magnetic fields on hollow cathode generated plasma columns in the APPEL device using both COMSOL Multiphysics fluid simulations and targeted experimental measurements. Our study focused on two working gases argon and helium to compare their plasma characteristics under magnetic fields ranging from 0 to 50 mT.

The axial plasma density profiles reveal that the application of an axial magnetic field significantly enhances plasma confinement and axial extension for both gases. In argon, increasing magnetic field strength transitions the plasma from a short, low-density plume at B = 0 mT to a highly collimated and elongated column with peak densities approaching ~ $6 \times 10^{17}$ m$^{-3}$. A saturation effect in axial expansion is observed beyond ~37 mT, indicating diminishing returns in further axial extension. Helium plasmas exhibit superior performance at comparable fields, with peak densities reaching ~ $4.5 \times 10^{18}$ m$^{-3}$ and a substantially more extended axial distribution attributable to helium's lower mass, higher thermal velocity, and enhanced electron mobility under magnetic confinement.

The electron temperature distributions demonstrate improved energy confinement with increasing magnetic field. In argon, the application of magnetic fields reduces peak electron temperature and spreads the energy distribution axially, indicative of reduced radial losses and stabilized confinement. Helium plasmas maintain higher and more uniform electron temperatures across the column, consistent with its higher ionization potential and reduced collisional loss mechanisms. These characteristics underscore helium's advantage in sustaining energetic electron populations over longer axial distances. Studies also reveals that the energetic electron population travels longer distance come in to center of plasma column and create the discharge current. A theoretical model for electron transport in a long magnetized plasma column was developed by formulating the electron continuity and momentum balance equations under magnetized conditions, where the cross-field diffusion is significantly suppressed and transport is dominated by motion along the magnetic field lines. In this framework, the perpendicular (cross-field) particle flux is expressed using classical and Bohm diffusion coefficients that depend on the magnetic field strength and plasma parameters, while the parallel (axial) flux follows from the field-aligned electron mobility and density gradients. This theoretical model establishes how magnetic confinement influences plasma column length and density distribution by reducing radial losses and enhancing axial electron transport.

Comparisons between experimental data and COMSOL simulations show good agreement in electron density and temperature at lower magnetic field values (0 to 20 mT). However, at higher fields (>30 mT), simulations tend to overestimate electron density relative to experimental measurements, likely due to unmolded transport losses, wall interactions, or plasma instabilities that are present in real discharges but not fully captured in the fluid model.

Analytical formulations based on cross-field diffusion balance and cylindrical conductor models were developed to estimate the magnetized plasma column length. These analytical predictions align qualitatively with COMSOL results, demonstrating increasing column length with magnetic field strength up to a saturation regime. For argon, all methods exhibit the expected trend of elongated plasma with magnetic field, with analytical estimates providing conservative bounds compared to simulation. For helium, analytical and simulation methods consistently predict substantial increases in plasma column length with field, though simulation results beyond 50 mT are limited.

Overall, this study demonstrates that axial magnetic fields play a crucial role in tailoring hollow cathode plasma characteristics by enhancing electron confinement, increasing plasma density, reducing radial diffusion, and extending the plasma column. Between the two gases studied, helium shows superior confinement and axial reach under identical field conditions, highlighting its promise for applications requiring long and dense plasma columns, such as beam neutralization, advanced materials processing, and electric propulsion. The combined use of experimental measurements, fluid simulations, and analytical modelling provides a comprehensive framework for understanding and optimizing magnetized hollow cathode discharges.

Although our analytical and numerical approaches both show that increasing the axial magnetic field leads to an extended plasma column, the absolute numerical values predicted for the helium plasma column length under high magnetic fields are unrealistically large when interpreted literally (e.g., tens of meters). Such magnitudes are not observed in laboratory magnetized plasma devices; for example, large linear plasma devices operating with background fields in the tens to a few hundred mT range typically produce plasma column lengths on the order of meters rather than tens of meters under steady conditions. The discrepancy arises because the simplified analytical model assumes idealized confinement and neglects important loss mechanisms such as enhanced plasma transport, wall interactions, finite neutral pressure effects, and end losses that become significant in real experimental systems.

## Acknowledgments

The authors would like to thank the Director of the Institute for Plasma Research (IPR). The authors acknowledge Magnetized Plasma Development section research scholars for their help and discussion.

## References

[1] Lieberman, M. A., & Lichtenberg, A. J. (2000). Radio-frequency plasma sources: fundamentals and applications. IEEE Transactions on Plasma Science, 28(1), 1–12. https://doi.org/10.1109/27.842325

[2] Chen, F. F. (1996). A detailed review on helicon plasma sources covering wave physics, plasma generation, and applications, Helicon Plasma Sources. IEEE Transactions on Plasma Science, 24(4), 1004–1012,

[3] Bacal, M. (2017). Electron Cyclotron Resonance Plasma Sources: Applications and Advances. Journal of Physics D: Applied Physics, 50(29), 293001.

[4] Kushner, M. J (1997), Modeling of hollow cathode discharges and the influence of geometry on plasma characteristics. Journal of Applied Physics, 82(5), 2471–2483. https://doi.org/10.1063/1.366344

[5] A. Anders, "Physics of Arc Cathodes," IEEE Trans. Plasma Sci., vol. 33, no. 2, pp. 145–157, 2005.

[6] J. W. Bradley et al., "Hollow cathode discharges and their applications," Plasma Sources Sci. Technol., vol. 13, no. 3, pp. 189–200, 2004

[7] Ahmed, M. A., Algwari, Q. T., & Younus, M. H. (2022). Plasma properties of a low-pressure hollow cathode DC discharge. International Journal of Recent Research and Review, 63(6), Article 20. https://doi.org/10.24996/ijs.2022.63.6.20

[8] Muhl, S., & Pérez, A. (2015). The use of hollow cathodes in deposition processes: A critical review. Thin Solid Films, 579, 174–198.

[9] M P Bhuva, S K Karkari and Sunil Kumar, Influence of cold hollow cathode geometry on the radial characteristics of downstream magnetized plasma column, Plasma Sources Sci. Technol., 28 (2019), 115013.

[10] G. Xia and G. Mao, "Performance prediction of a novel microplasma thruster with microhollow cathode discharge," Bulletin of the American Physical Society, vol. 54, no. 12, Abstract KTP.00094, 2009.

[11] Hassouba, M. A. (2001). Effect of the magnetic field on the plasma parameters in the cathode fall region of the DC-glow discharge. The European Physical Journal Applied Physics, 14(2), 131–135.

[12] J. H. Kim, "Development of the DC-RF hybrid plasma source and its application to the etching and texturing of the silicon surface," Ph.D. dissertation, Jeju National University, Jeju, Korea, 2011.

[13] Ness, H., & Makabe, T. (2000). Behavior of charged particles in electric and magnetic fields in low-temperature plasma processing devices. Journal of Vacuum Science & Technology A, 18(4), 1567–1573.

[14] Y. Patil, S. Karkari, Fluid simulation studies of low temperature plasmas using COMSOL Multiphysics Software, 8[th] Asia-Pacific Conference on Plasma Physics, 3-8 Nov, 2024 at Malacca, https://www.aappsdpp.org/DPP2024/html/3contents/pdf/5622.pdf.

[15] S. Binwal, Y. Patil, S. K. Karkari, and L. Nair, Transverse magnetic field effects on spatial electron temperature distribution in a 13.56 MHz parallel plate capacitive discharge, Phys. Plasmas 27(3), 033506 (2020).

[16] Y. Patil, S K Karkari, M A Ansari, Varun Dixit, Ravi Ranjan, Raj Singh, Dhyey Raval, P K Sharma and Raju Daniel, APPEL device experimental studies on spiral antenna for efficient preionization in SST-1 tokamak, Plasma Phys. Control. Fusion 67 (2025) 075026.

[17] P. Singh, A. Pandey, S. Dahiya, Y. Patil, N. Sirse, and S. Karkari, Sheath effects with thermal electrons on the resonance frequency of a DC-biased hairpin probe, Physics of Plasmas, vol. 32, no. 10, 103511, 2025.

[18] A. Khandelwal, D. Raval, N. Sharma, Y. Patil, S. Sharma, S. Karkari, and N. Sirse, Experimental Studies of an Axially Symmetric Magnetized Cylindrical Capacitively Coupled Plasma Discharge, https://ieeexplore.ieee.org/document/11261847.

[19] S. Dahiya, P. Singh, Y. Patil, S. Sharma, N. Sirse, and S. Karkari, Discharge characteristics of a low-pressure geometrically asymmetric cylindrical CCP with an axisymmetric magnetic field, Phys. Plasmas 30, 093505 (2023).

[20] Patil, Y., & Karkari, S. (2023). Applied Plasma Physics Experiments in Linear (APPEL) device for plasma surface interaction studies. Fusion Engineering and Design, 197, 114056. https://doi.org/10.1016/j.fusengdes.2023.114056.

[21] Adamovich, I., Agarwal, S., Ahedo, E., Alves, L. L., Baalrud, S., Babaeva, N., Bogaerts, A., Bourdon, A., Bruggeman, P. J., Canal, C., Choi, E. H., Coulombe, S., Donkó, Z., Graves, D. B., Hamaguchi, S., Hegemann, D., Hori, M., Kim, H., Kroesen, G. M. W.,Von Woedtke, T. (2022). The 2022 Plasma Roadmap: low temperature plasma science and technology. Journal of Physics D Applied Physics, 55(37), 373001. https://doi.org/10.1088/1361-6463/ac5e1c

[22] I. L. Alberts, D. S. Barratt, and A. K. Ray, Hollow Cathode Effect in Cold Cathode Fluorescent Lamps: A Review, J. Disp. Technol., vol. 6, no. 2, pp. 52–59, Feb. 2010, doi: 10.1109/JDT.2009.2031924.

[23] Y. Patil, S. Binwal, and S. K. Karkari, Fluid modeling of E × B drift occurs in magnetized CCP discharge using comsol multiphysics, in 12th International Conference on Plasma Science and Applications (ICPSA-2019), 2019.

[24] Cheng Jia et al, Fluid model of inductively coupled plasma etcher based on COMSOL, J. Semicond., 2010, 31 032004.

[25] Lymberopoulos D P, Economou D J, 2-dimensional self-consistent radio-frequency plasma simulations relevant to the gaseous electronics conference RF reference cell. J Res Natl Inst Stand Technol, 1995, 100(4): 473.

[26] Yinghua Liu et al, Numerical simulation of dynamics behavior of pulsed-DC helium plasma jet confined by parallel magnetic field at atmospheric pressure, Phys. Rev. Research 6, 033028.

[27] http://jila.colorado.edu/~avp/collision_data/electronneutral/ELECTRON.TXT.

[28] F. Iza, S. H. Lee, and J. K. Lee, "Computer modeling of low-temperature plasmas," in Gas Discharges: Fundamentals & Applications, Thin Films and Nanostructures, edited by J. A. Filho (Transworld Research Work, 2007), pp. 1–31.

[29] K. Ohya, T. Ishitani, Target material dependence of secondary electron image induced by focused ion beams, Surf. Coat. Technolo. 158-159 (2002) 8-13.

[30] M. A. Lieberman and A. J. Lichtenberg, *Principles of Plasma Discharges and Materials Processing*, 2nd ed., Wiley-Interscience, Hoboken, NJ, 2005.

[31] Bhuva M P, Karkari S K and Kumar S, Characteristics of an elongated plasma column produced by magnetically coupled hollow cathode plasma source Phys. Plasmas, 2018 25 033509.